\documentstyle[prb,aps,epsf]{revtex}

\draft
\newcommand{\beq}{\begin{equation}}
\newcommand{\eeq}{\end{equation}}

\begin{document}

\title{Barkhausen noise in soft amorphous magnetic
materials under applied stress}

\author{Gianfranco Durin$^1$ and Stefano Zapperi$^{2}$}

\address{$^1$ Istituto Elettrotecnico Nazionale Galileo Ferraris and
        INFM, Corso M. d'Azeglio 42, I-10125 Torino, Italy\\
        $^2$ PMMH ESPCI, 10 rue Vauquelin, 75231 Paris-Cedex 05,
        France}

\maketitle

\begin{abstract}
We report experimental measurements of Barkhausen noise on
Fe$_{64}$Co$_{21}$B$_{15}$ amorphous alloy
under tensile stress. We interpret the scaling behavior of the noise
distributions in terms of the depinning
transition of the domain walls. We show that stress induced
anisotropy enhance the effect of short-range
elastic interactions that dominate over long-range dipolar
interactions. The universality class is thus different
from the one usually observed in Barkhausen noise measurements
and is characterized by the exponents $\tau\simeq 1.3$ and
$\alpha\simeq 1.5$, for the decay of the distributions of jump sizes
and durations.
\end{abstract}


\section{Introduction}

The Barkhausen effect has received a renewed theoretical interest
in the last few years, because of its connections with interface
depinning and non-equilibrium critical phenomena. We have recently
proposed \cite{czds,zcds} a microscopic model of the Barkhausen
effect based on the dynamics a flexible domain wall on a random
environment. The model gives a satisfactory explanation of many of
the experimental results reported in literature. In particular, we
are able to justify the occurrence of the power law distributions
of Barkhausen jumps and compute the critical exponents. In the
limit of a vanishing field driving rate, the size distribution
scales as $P(s)\sim s^{-\tau}f(s/s_0)$, where $\tau = 3/2$, and
the cutoff scales with the demagnetizing factor as $s_0\sim 1/k$.
Similarly, the duration distribution scales as $P(T) \sim
T^{-\alpha}g(T/T_0)$, where $\alpha=2$, and the cutoff scales as
$T_0\sim 1/\sqrt{k}$. These results are the strict consequence of
the presence of long-range dipolar interactions and are in perfect
agreement with experiments on amorphous alloys and polycrystalline
samples \cite{fractal,durin4,mamexp}.

In a previously introduced model \cite{urbach}, based on a
description of the domain wall similar to ours, dipolar
interactions were ignored, giving rise to different critical
exponents $\tau \sim 1.3$, and $\alpha \sim 1.5$. Comparable
experimental values have only been reported for Perminvar (30\%
Fe, 45\% Ni, 25\% Co) \cite{urbach}, where $\tau\sim 1.33 $. In
this paper, we show that the critical exponents in
Fe$_{64}$Co$_{21}$B$_{15}$ amorphous alloy under tensile stress
display the values $\tau \sim 1.28$, and $\alpha \sim 1.5$,
indicating that the experiment is described by short-range
interface models. We explain these results noting that for
magnetostrictive materials, such as Fe$_{64}$Co$_{21}$B$_{15}$,
magneto-elastic interactions enhance the effect of the surface
tension over the long-range dipolar forces.

\section{Theory}

Before describing in detail the experimental results, we summarize
the main theoretical results recently discussed in
Refs.~\onlinecite{czds,zcds}. We consider a single $180^{\circ}$
domain wall, described by its position $h(\vec{r})$, dividing two
regions of opposite magnetization directed along the $x$ axis. The
total energy for a given configuration is the sum of different
contributions due to magnetostatic, ferromagnetic and
magneto-crystalline interactions, and gives rise to the following
equation of motion \cite{czds,zcds} \beq \frac{\partial
h(\vec{r},t)}{\partial t}= H-k \bar{h}+ \nu_0 \nabla^2h(\vec{r},t)
+ \int d^2r^\prime K(\vec{r}-\vec{r}^{\;\prime})(h(\vec{r
}^{\;\prime})-h(\vec{r})) +\eta(\vec{r},h), \label{eqm} \eeq where
$H$ is the applied field, $k$ is the demagnetizing factor and
$\bar{h} \equiv \int d^2r^\prime h(\vec{r}^{\;\prime},t)/V$,
$\nu_{0}$ is the surface tension of the wall, $K$ is the kernel
due to dipolar interactions given by \beq
K(\vec{r}-\vec{r}^{\;\prime})=
\frac{\mu_0M_s^2}{2\pi|\vec{r}-\vec{r}^{\;\prime}|^3}\left(1+
\frac{3(x-x^\prime)^2}{|\vec{r}-\vec{r}^{\;\prime}|^2}\right),
\label{eq:ker} \eeq and $\eta(\vec{r},h)$ is a Gaussian
uncorrelated random field taking into account all the possible
effects of dislocations, residual stress and non-magnetic
inclusions.

When $k=0$, Eq.~(\ref{eqm}) displays a depinning transition as a function
of the field and the corresponding critical exponents
can be obtained with renormalization group methods
\cite{natt,nf}. For $k>0$, which corresponds to the experiments
we are describing here, the domain wall displays a steady
motion {\it around} the depinning transition \cite{urbach}.
For low driving rates, the velocity of the wall shows avalanches
whose distributions are characterized by the scaling laws
expected close to the depinning transition.

The upper critical dimension of Eq.~(\ref{eqm}) is $d_c=3$
 so that the dynamics can be described by a mean-field
equation\cite{czds}. This is the reason of the success of a mean-field
phenomenological model \cite{abbm} to explain most
of the experimental data. In fact, the critical exponents take
mean-field values $\tau = 3/2$, and $\alpha=2$, with a
small correction at non-zero driving rates, $\tau = 3/2 -c/2$, and
$\alpha=2-c$, where $c$ is a dimensionless parameter proportional
to the driving rate $dH/dt$, in agreement with experimental data
on many different materials \cite{zcds,fractal,durin4,mamexp}.

These results change when the long-range kernel
(Eq.~(\ref{eq:ker})) is absent and the surface tension is the most
relevant term in the equation. The upper critical dimension
changes to $d_c=5$, with critical exponents significantly lower,
$\tau = 1.25$, and $\alpha=1.43$ from renormalization group
calculations\cite{natt,nf,heiko}, and $\tau \simeq 1.3$, and
$\alpha \simeq 1.5$ from simulations \cite{heiko,koiler}.

\section{Experimental results and simulations}

Ribbons of an high magnetostrictive
($\lambda_{s}=46.5\cdot10^{-6})$ Fe$_{64}$Co$_{21}$B$_{15}$
amorphous alloy are prepared by planar flow casting in air. We
employ a single strip of 30 cm length under tensile stress from 0
to 140 MPa. The applied stress induces uniaxial anisotropies,
radically changing the domain structure. This kind of high
magnetostrictive alloy shows a typical behavior of the coercive
field $H_{c}$ (see for instance \cite{appino}): a fast decrease of
$H_c$ up to a typical stress value $\sigma_{0} \sim 10 MPa $, and
a roughly linear increase at high stresses. This behavior is due
to changes in the domain structure: the quenched-in stresses
dominate at low tensile stress, resulting in a complicated pattern
of maze domains. As the external stress increases, and the
uniaxial anisotropy is induced, a simpler domain structure
appears, with a few parallel domains in the direction of the
stress \cite{appino}.

\begin{figure}[th]
\widetext \centerline{
        \epsfxsize=10.5cm
        \epsfbox{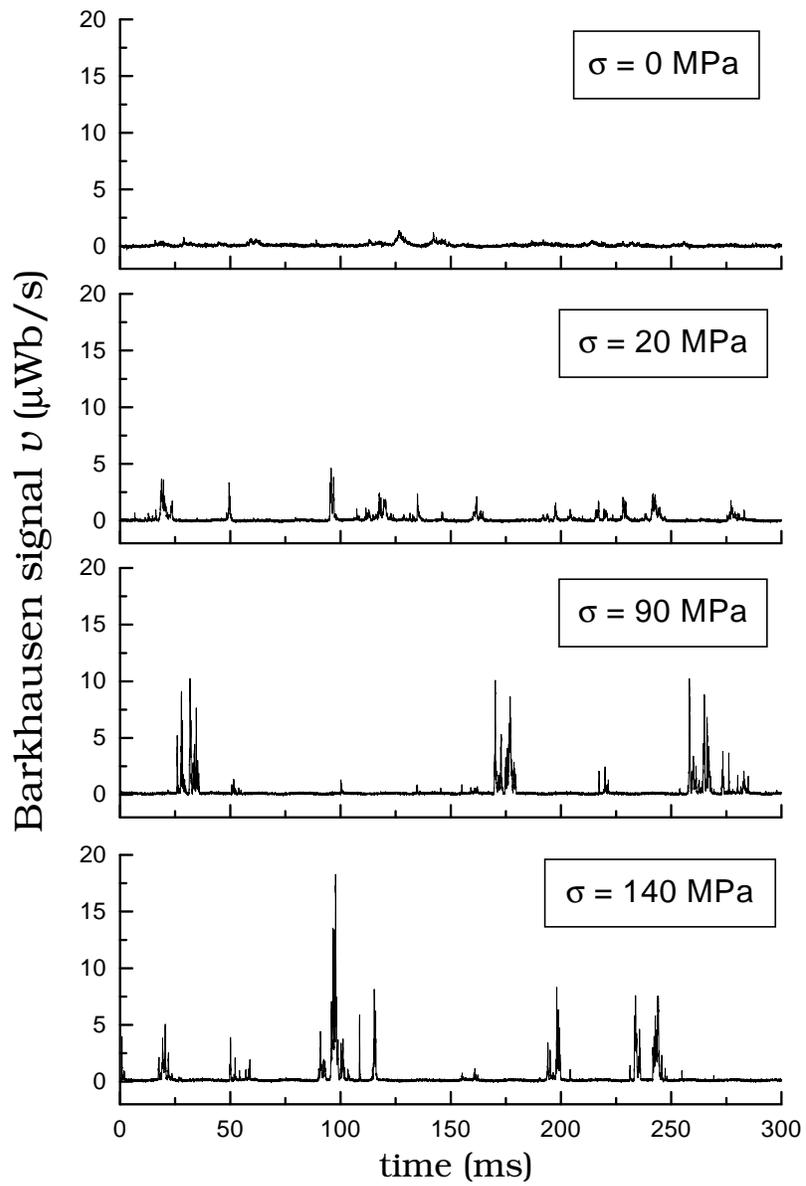}
        \vspace*{1cm}
        }
\caption{Barkhausen signal $v$ at different tensile stresses for
the as-cast Fe$_{64}$Co$_{21}$B$_{15}$ amorphous
alloy.}\label{signal}
\end{figure}

The Barkhausen noise measurements are performed as explained in
Ref.~\onlinecite{fractal}. The visual inspection of Barkhausen
signal reveals the changes induced by the stress in the domain
wall dynamics (Fig.~\ref{signal}). The amplitude of the signal
increases with the stress, resulting in very short and peaked
jumps at high stresses. A better characterization of the
Barkhausen jump shape is obtained considering the distributions of
the amplitude $v$, the size $s$, and the durations $T$
\cite{fractal,bdm}.

\begin{figure}[h]
\widetext \centerline{
        \epsfxsize=10.5cm
        \epsfbox{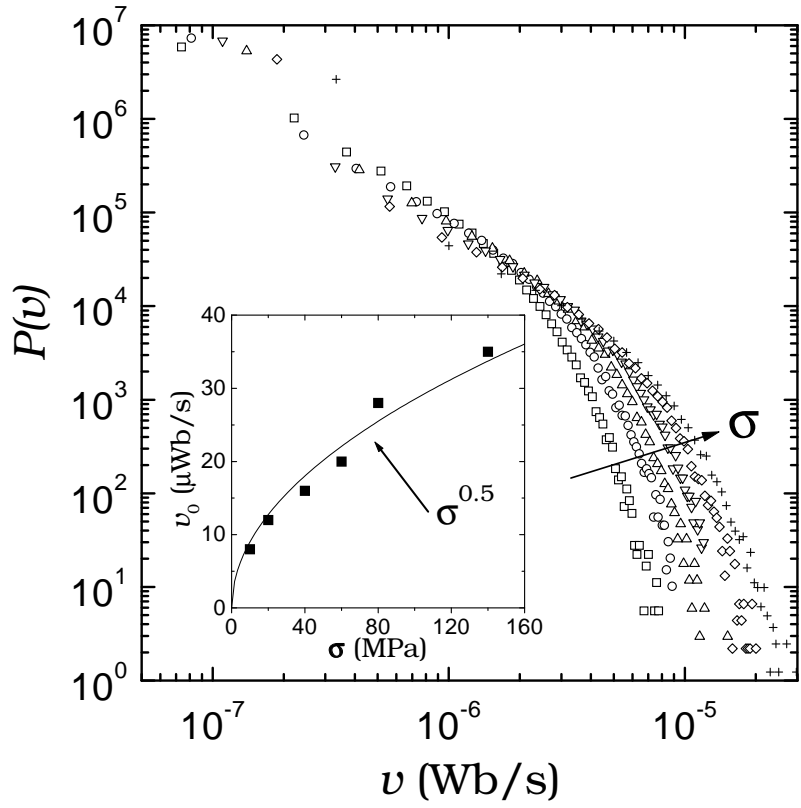}
        \vspace*{1cm}
        }
\caption{Distribution of the Barkhausen signal amplitude $v$ at
different tensile stresses ($\sigma =$ 10-140 MPa). Data follows
the distribution $P(v) \sim v^{-(1-c)}exp(-v/v_0)$, where $c$ is
the normalized driving rate. (Inset) Dependence of the cutoff
$v_0$ on the applied stress, showing $v_0 \sim
\sigma^{0.5}$.}\label{Pv}
\end{figure}

\begin{figure}[th]
\widetext\centerline{
        \epsfxsize=10.5cm
        \epsfbox{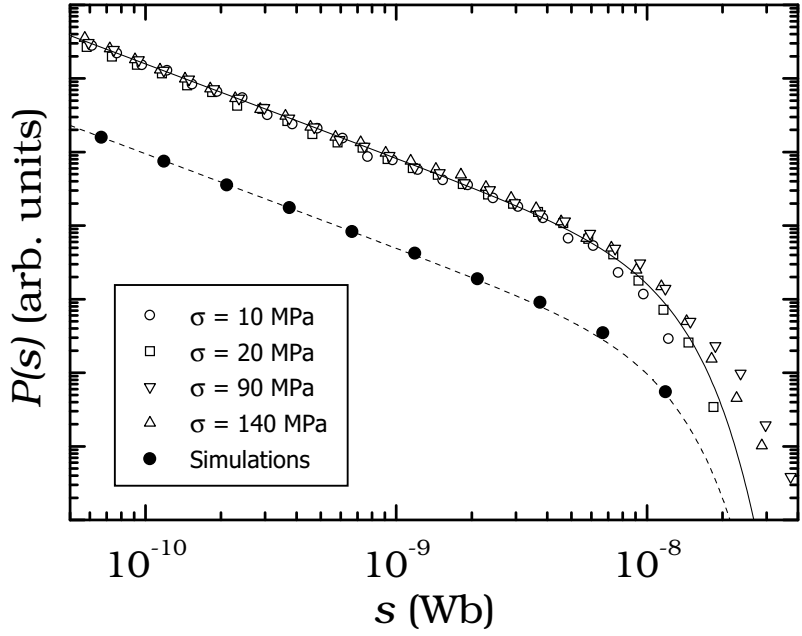}
        \vspace*{1cm}
        }
\caption{Experimental size distributions of Barkhausen jumps at
different tensile stresses (open dots), compared to the data
(solid dots) simulated using Eq.~(\ref{eqm}) without dipolar
interactions. All data are fitted using $P(s) \sim
s^{-\tau}exp(-(s/s_0)^2)$, where $\tau = 1.26$ (experiments), and
$\tau = 1.28$ (simulations), with $s_o \sim 10^{-8}$ Wb. Simulated
data are rescaled for comparison.}\label{Ps}
\end{figure}

As in previous experiments \cite{fractal,bdm,smm}, the amplitude
distribution is found to decay as \beq P(v) \sim
v^{-(1-c)}exp(-v/v_0), \eeq where $c$ is the normalized driving
rate. We show in Fig.~\ref{Pv} that the cutoff $v_0$ increases
with the stress, scaling as $v_0\sim\sigma^{0.5}$. Size and
duration distributions are collected at nearly same the value of
$c = 0.1 - 0.2$, showing exponents $\tau \sim 1.28\pm 0.02$
(Fig.~\ref{Ps}), and $\alpha \sim 1.5\pm 0.1$, {\it independent}
of the applied stress, provided that $\sigma > \sigma_{0}$. At low
applied stress ($\sigma < \sigma_{0}$), where the domain structure
is more complicated and the Barkhausen signal/background noise
ratio is reduced, it is difficult to obtain reliable estimates of
the exponents.

For $\sigma \simeq \sigma_{0}$, we also measured the variations of
the critical exponents with the parameter $c$. Unlike mean-field
theory \cite{abbm}, here the dependence from $c$ is quite weak
within the fitting error bar, in agreement with the result
reported for the Perminvar \cite{urbach}.

The cutoff of the distributions shows an intriguing behavior: the
cutoff of the size distribution $s_0$ is nearly independent on the
stress (Fig.~\ref{Ps}), while the duration is progressively
reduced with a stress dependence of the type $T_{0} \sim
\sigma^{-0.5}$. These results quantify the qualitative picture of
the signal discussed above: the amplitude of the signal increases
with the stress, while avalanche durations are shorter, so that
the avalanche sizes roughly do not change with the stress.

In order to understand the experimental results, we perform a serie of
simulations of a discretized version of Eq.~(\ref{eqm}) without
the long-range kernel \cite{urbach}. We employ different field
driving rates and evaluate the parameter $c$ using
the amplitude distributions. We observe small variations in the
critical exponents: $\tau = 1.26\pm0.04 - (0.16\pm 0.05)c$ and
$\alpha = 1.40\pm 0.05 - (0.25\pm 0.05) c$, in excellent
agreement with experiments.

\section{Discussion}

The critical exponents of the experimental Barkhausen signal in
a magnetostrictive sample under stress are
in agreement with the results of our simulations and those previously
reported \cite{urbach,koiler}, neglecting dipolar interaction
in the equation of motion. This implies that simple domain
structures with a few straight domains, spanning from end to end,
as those induced by tensile stress, are well described by the
universality class of the equation with elastic surface tension.
This observation provides an interesting experimental case where
accurate results can be compared with statistical theories of
interface depinning.

In general terms, if we believe that Eq.~(\ref{eqm}) describes
all the essential elements of the domain wall dynamics, we
expect that the experimental results will belong to either of the
two universality classes considered above.  In general, it is not
easy to predict the universality class from the
knowledge of the domain structure of a sample,
and its magnetic or thermal treatment.
We can, however, estimate the relative weights of the dipolar
and elastic energies and obtain a rough classification of the
results that can be observed in different materials.

Roughly speaking, the energy associated to the ``elastic'' tension
of the wall is $E_{el} \sim\nu_{0}S$, where $S$ is the area of
domain wall, while the magnetostatic energy is of the order of
$E_{ms}\sim\mu_{0}M_{s}^{2}k^{*}V$, where $k^{*}$ is the
demagnetizing factor associated with the region of volume $V$
where magnetic charges appear \cite{Libro}. As a first
approximation, we can take $V = S \delta_{w}$, where $\delta_{w}$
is the domain wall width. Since $\nu_{0} \sim \sqrt{AK}$, and
$\delta_{w} \sim \sqrt{A/K}$, where $A$ is the exchange constant
and $K$ the anisotropy constant, the ratio between the two terms
is of the order of $r\equiv E_{el}/E_{ms}\sim 10^{-6}K/k^{*}$,
since $\mu_o M_s \sim 1 T$. The factor $k^{*}$ cannot be easily
evaluated, because it depends strongly on the spatial distribution
of the charges. It vanishes for a perfectly rigid wall and is of
order 1 for a large bending of the wall. In general, considering
only a small bending of the surface, we can estimate $k^* \sim
10^{-2}$, which correspond to length/width ratio of the order of
10 or more, giving $r \sim 10^{-4}K$. In soft amorphous alloys as
the one we employ in our experiments, the average anisotropy
constant $\langle K \rangle$ is of the order of 200 J/m$^3$
(Ref.~\onlinecite{appino}), so that $r \ll 1$, and magnetostatic
effects prevail \cite{zcds,smm}. Under applied stress, the induced
anisotropy ($K_\sigma\sim 3/2\lambda_{s}\sigma$), and the reduced
value of $k^{*}$, yield $r\geq 1$, so that the surface tension
becomes dominant. A similar effect could also explain the
experimental results on Perminvar \cite{urbach}, a material with a
low permeability value ($\sim 250$) consistent with a quite high
value of $K \sim10^{4}- 10^{5}$.

In principle, it should be possible to observe the transition between the
two universality classes changing the induced anisotropy in an
amorphous alloy. Our results do not definitely prove it, since the
exponents at low stress have quite large error bars. Experiments
on other amorphous alloys are in progress to confirm this
hypothesis. Finally, we are currently working to understand
quantitatively the scaling of the cutoff of the distributions
with the applied stress.

\newpage

\newpage

\end{document}